\documentclass{elsart}
\usepackage{amsmath,amssymb,eurosym,graphicx,rotating,threeparttable,url}
\usepackage{apacite}
\renewcommand{\BBAA}{and}
\usepackage[sort]{natbib}
\journal{JEIC}
\begin{document}


\begin{frontmatter}
\title{Social networks and labour productivity in Europe: An empirical investigation}
\author[Ancona,Canberra]{C. Di Guilmi\corauthref{cor}},
\corauth[cor]{Corresponding author. Tel.: +39--071--22--07--112; fax: +39--071--22--07--102.}
\ead{c.diguilmi@univpm.it}
\author[Ancona,Canberra]{F. Clementi},
\ead{fabio.clementi@univpm.it}
\author[Canberra]{T. Di Matteo},
\ead{tiziana.dimatteo@anu.edu.au}
\author[Ancona]{M. Gallegati}
\ead{mauro.gallegati@univpm.it}
\address[Ancona]{Department of Economics, Polytechnic University of Marche, Piazzale R. Martelli 8, 60121 Ancona, Italy}
\address[Canberra]{Applied Mathematics, Research School of Physical Sciences and Engineering, The Australian National University, 0200 Canberra, Australia}


\begin{abstract}
This paper uses firm-level data recorded in the \textsc{Amadeus} database to investigate the distribution of labour productivity in different European countries. We find that the upper tail of the empirical productivity distributions follows a decaying power-law, whose exponent $\alpha$ is obtained by a semi-parametric estimation technique recently developed by \citet{ClementiDiMatteoGallegati2006}. The emergence of ``fat tails'' in productivity distribution has already been detected in \citet{DiMatteoAsteGallegati2005} and explained by means of a model of social network. Here we show that this model is tested on a broader sample of countries having different patterns of social network structure. These different social attitudes, measured using a social capital indicator, reflect in the power-law exponent estimates, verifying in this way the existence of linkages among firms' productivity performance and social network.
\end{abstract}
\begin{keyword}
Labour productivity \sep power-law distribution \sep semi-parametric bootstrap approach \sep social networks \sep social capital
\PACS 02.50.Tt\sep 89.65.-s\sep 89.75.Fb
\end{keyword}
\end{frontmatter}


\section{Introduction}
A consistent flow of research on firms' and workers' productivity regarding the topic of technology innovation and diffusion has focused on generation and transmission of innovations through networks of firms \citep{PittawayRobertsonMunirDenyerNeely2004}, or on the relationship among social capital and productivity performance \citep{CohenPrusak2001}\footnote{See \citet{Rogers2003} for a comprehensive topic review.}. As highlighted in \citet{DiMatteoAsteGallegati2005}, firms' network plays a decisive role in the imitation process of the innovative firms through which, according to the evolutionary literature perspective, innovations originally conceived by a given firm \textit{percolate} outside it by imitation from other firms. In this way the innovation flows through the network of contacts and communications between firms. The significance of the underlying connection network comes into sight when the collective dynamics of the system is considered. As showed in several studies, above a certain threshold of complexity, natural, artificial, and social systems are typically characterized by networks with power-law degree distribution, \textit{i.e.} ``scale-free'' networks \citetext{see \textit{e.g.} \citealp{AlbertBarabasi2002}}. In very recent times, network theory gained \textit{momentum} in explaining firms' performance also from a technical perspective. The amazingly rapid progress that took place in information technologies since the mid of '90s accounts for a noteworthy proportion of productivity growth. Contemporarily, it also broadened the role of networks in determining firms' labour productivity performances. The conjunct use of information networks along supply or customer chains pushed toward a higher specialization and improvement of skills in labour force and, in general, leaded to remarkable changes in the competences needed within firms in order to maintain competitiveness on the market \citep{Motohashi2007}.
\par
In this paper we extend the analysis of the relationship among network and productivity in two directions. First, we exploit the link between social capital, social network and productivity distribution among firms. We do not limit our analysis to the firms' network \citetext{see \textit{e.g.} \citealp{Ahuja2000}}, but we \textit{embed} it in the study of social network characteristics, treating therefore also the non-economic aspects that determine the social environment in which firms operate and interact. According to \citet{Granovetter2005}, the social network influences firms' productivity through different channels: the mutual acceptance and the prizing of technical skills inside the community of workers within a firm; the control exerted among colleagues, that determines the quality of the effort and, therefore, the efficiency of single workers in a way analogous to principal-agent models; the interpersonal ties, inside and outside the firms, enforced by repeated interaction, that lead to a level of trust that eases the interrelations and the flow of information.
\par
The second aspect of novelty consists in the method of analysis. Indeed, the impact of social network structure on productivity is quantitatively evaluated by means of labour productivity distribution features, in order to verify whether and to which extent social systems and social capital favor the circulation of information and innovation through networks of firms. The differences recorded among firms' productivity levels within a country determine the shape of productivity distribution. As evidenced by \citet{Coleman1988}, stronger network ties make the circulation of information faster and less expensive. This, in turn, may reduce the gap in performances across firms by favoring the transmission of knowledge and innovations, and thus leading to a more even distribution of productivity among firms. Therefore, in this paper we investigate how differences in social capital reflect into disparities in productivity distribution shapes and parameters.
\par
The study proceeds as follows: in Section \ref{sec:PowerLawDecayInProductivityDistribution}, it is examined whether labour productivity follows a power-law distribution in a sample of 9 European countries. This assessment is of particular interest, since the sample of countries in object are not homogeneous from both an economic and a social point of view. The presence of power-law tails in such different contexts might reveal that this emergence does not depend on a particular underlying social structure, but it is consistent over different systems. The estimates of the power-law exponent are here obtained by means of the technique introduced by \citet{ClementiDiMatteoGallegati2006}. This method adopts a subsample semi-parametric bootstrap algorithm for optimally selecting the number of extreme quantiles to be used in the upper tail estimation, and thus ending up with less ambiguous estimates of the exponent $\alpha$. Furthermore, we model the network of firms along the lines of \citet[but see also \citealp{DiMatteoAsteHyde2004}]{DiMatteoAsteGallegati2005}: the use of this model allows to get a quantitative measure of the role of the underlying network of firms in determining the shape of productivity distribution. According to this work, the emergence of ``fat-tailed'' distributions may be interpreted as the outcome of an analogous structure of the network, which must show slow decaying tails in its degree distribution, and, therefore, a ``scale-free'' type behaviour\footnote{\citet{PammolliRiccaboni2001} sustain this interpretation by detecting power-law distributions in firms' networks.}. In Section \ref{sec:SocialNetworkSocialCapitalAndEconomicPerformance}, the link between networks of firms and social networks is illustrated by comparing the tail exponents of the labour productivity distributions to a social capital indicator by country, and also testing if social capital influences the aggregate growth of productivity. Finally, Section \ref{sec:Conclusions} summarizes and concludes.


\section{Power-law decay in productivity distribution}
\label{sec:PowerLawDecayInProductivityDistribution}
Our aim here is to perform tail parameter estimations on labour productivity data through a recently developed method. The labour productivity is defined as added value over the amount of employees (where added value, defined according to standard balance sheet reporting, is the difference between total revenue and cost of inputs excluding the cost of labour). The results are used in the remainder of this section to link our empirical findings to a model of firms' interaction across a complex network.


\subsection{Data and methodology}
In this paper we have used the \textsc{Amadeus} database, compiled by Bureau van Dijk Electronic Publishing\footnote{Further details on the database can be found on the provider website: \url{http://www.bvdep.com/en/amadeus.html}.}. This data source contains firm-level data from all over Europe, and is available in different sizes. Firms in this study are taken from the ``TOP 250,000 Module'', including companies that fulfill one of three criteria regarding the magnitude of operating revenues, total assets and the number of employees\footnote{For France, Germany, Italy, Russia, Spain, the United Kingdom and Ukraine, the inclusion thresholds are \EUR{15} million in operating revenues, \EUR{30} million in assets and 200 employees. For all the other countries, they are \EUR{10} million in operating revenues, \EUR{20} million in assets and 150 employees.}. The analysis is based on 10 years of data (1996--2005) for 9 countries (Belgium, Finland, France, Germany, Italy, Netherlands, Spain, Sweden and the United Kingdom); for some of them (Germany and Italy) we have also used data by geographical sub-areas (East/West Germany and North/South Italy, respectively). The number of observations for each year and country is shown in Table \ref{tab:Table1}.
\begin{sidewaystable}[p]
\centering
\caption{The number of companies from 1996 to 2005 on \textsc{Amadeus} database (Bureau Van Dijk) in the following countries and geographical sub-areas: Belgium (BEL), Finland (FIN), France (FRA), Germany (GER), East Germany (EASTGER), West Germany (WESTGER), Italy (ITA), North Italy (NORTHITA), South Italy (SOUTHITA), Netherlands (NET), Spain (SPA), Sweden (SWE) and the United Kingdom (UK).}
\label{tab:Table1}
\vspace{0.5cm}
\begin{threeparttable}
\begin{tabular}{c|c|c|c|c|c|c|c|c|c|c|}
\cline{2-11}
&1996&1997&1998&1999&2000&2001&2002&2003&2004&2005\\
\hline
\multicolumn{1}{|c|}{BEL}&4,205&4,396&4,733&5,076&5,378&5,698&6,104&6,240&6,258&1,314\\
\multicolumn{1}{|c|}{FIN}&981&1,432&1,416&1,752&1,901&2,030&2,283&2,470&2,451&1,255\\
\multicolumn{1}{|c|}{FRA}&9,239&10,745&12,450&13,045&13,570&14,204&15,468&15,911&17,855&3,037\\
\multicolumn{1}{|c|}{GER}&1,453&1,497&1,656&1,661&1,961&2,126&3,807&4,404&4,278&751\\
\multicolumn{1}{|c|}{EASTGER}&186&192&233&246&274&269&512&583&599&107\\
\multicolumn{1}{|c|}{WESTGER}&1,267&1,305&1,423&1,415&1,687&1,857&3,295&3,821&3,679&644\\
\multicolumn{1}{|c|}{ITA}&10,904&11,861&12,087&13,742&14,360&14,995&16,492&13,574&16,715&3,586\\
\multicolumn{1}{|c|}{NORTHITA}&7,945&8,604&8,845&9,956&10,312&10,682&11,834&10,292&12,037&3,011\\
\multicolumn{1}{|c|}{SOUTHITA}&2,949&3,257&3,242&3,786&4,048&4,313&4,658&3,282&4,678&575\\
\multicolumn{1}{|c|}{NET}&1,404&1,643&1,884&2,685&2,616&3,221&3,961&4,112&3,825&807\\
\multicolumn{1}{|c|}{SPA}&6,551&7,382&8,356&9,020&10,123&11,378&12,472&12,736&12,300&228\\
\multicolumn{1}{|c|}{SWE}&n.a.\tnote{\textit{a}}&2,437&3,674&5,815&6,387&6,855&7,278&7,517&7,728&2,768\\
\multicolumn{1}{|c|}{UK}&4,205&10,563&11,578&12,545&13,679&15,082&16,482&17,342&17,687&5,996\\
\hline
\end{tabular}
\begin{tablenotes}
\item[\textit{a}]\footnotesize n.a. = Data not available.
\end{tablenotes}
\end{threeparttable}
\end{sidewaystable}
It should be noted that the number of companies for all countries is lower in 1996 and 2005 compared to all other years in the time span. Therefore, results from these years should be used with caution, since they might not be completely reliable.
\par
From these data we have calculated the empirical complementary cumulative distributions ($P_{\geq}\left(x\right)$, being the probability to find a firm with productivity larger than or equal to $x$), which show a very clear linear trend for large values of $x$ in a log-log scale, implying a non-Gaussian character with the probability for large productivities well described by a power-law behaviour, \textit{i.e.} $P_{\geq}\left(x\right)\sim x^{-\alpha}$. To extract the value of $\alpha$ we have used \citeauthor{ClementiDiMatteoGallegati2006}'s \citeyearpar{ClementiDiMatteoGallegati2006} subsample semi-parametric bootstrap algorithm for \textit{data-driven} selection of the number of observations located in the tail of the distribution. This technique relies on the popular \citeauthor{Hill1975}'s \citeyearpar{Hill1975} maximum likelihood estimator for the tail index $\alpha$, given by
\begin{equation}
\alpha_{n}=\left\{\frac{1}{m}\sum\limits^{m}_{i=1}\left[\log x_{\left(n-i+1\right)}-\log x_{\left(n-m\right)}\right]\right\}^{-1},
\label{eq:Equation1}
\end{equation}
where $n$ is the sample size, $m$ the number of observations in the tail of the distribution and the sample elements are put in descending order, \textit{i.e.} $x_{\left(n\right)}\geq x_{\left(n-1\right)}\geq\cdots\geq x_{\left(n-m\right)}\geq\cdots\geq x_{\left(1\right)}$. As well known, the main problem connected with the Hill's estimator is the decision about an appropriate tail size, \textit{i.e.} the optimal number of observations $m$ included in the calculation of $\alpha_{n}$. This choice is accomplished by the authors through minimisation of the finite-sample Mean Squared Error (MSE) of the estimator \eqref{eq:Equation1}, so that an optimal $m$ is defined by
\[m^{\ast}=\arg\min\limits_{m}E\left[\left(\alpha_{n_{1}}^{\#}-\alpha_{n}\right)^{2}\right],\]
where $\alpha_{n}$ is an initial estimate from the original sample and $\alpha_{n_{1}}^{\#}$ is the estimate obtained using the bootstrapped datasets drawn from a smoothed parametric distribution of $n_{1}\leq n$ observations belonging to the null hypothesis of a complete sequence of goodness-of-fit tests for Pareto-type tail behaviour. The number of bootstrap replications is automatically chosen according to a three-step procedure to achieve the desired level of accuracy, where accuracy is measured by the percentage deviation of the estimate obtained by running a finite number of bootstrap repetitions from the corresponding ideal bootstrap quantity estimated with an infinite number of resamples \citep{AndrewsBuchinsky2000}. Since MSE comprises the variance and bias of the estimator, the optimal estimate $\alpha_{n}^{\ast}$\textemdash making use of $m^{\ast}$ observations lying in the tail\textemdash will be in this way a balance between the former (which usually decreases with increasing tail size) and the latter (which tends to increase with tail size)\footnote{Hill himself devised a data-analytic method for choosing $m^{\ast}$ which is based on sequentially testing appropriate functions of the observations for exponentiality. However, the application of this procedure to our productivity data resulted in overestimation of the tail exponent compared to the semi-parametric bootstrap algorithm. This appears to empirically support \citeauthor{HallWelsh1985}'s \citeyearpar{HallWelsh1985} argument of a very gradual deterioration of the exponential approximation, leading Hill's method to largely overestimate $m$ (and thus $\alpha$ by Eq. \eqref{eq:Equation1}).}.
\begin{sidewaystable}[p]
\centering
\caption{Estimates of the power-law tail exponent $\alpha$ by subsample semi-parametric resampling and 95\% confidence intervals\textit{\textsuperscript{a}}. The data used are taken from the \textsc{Amadeus} database by Bureau Van Dijk, and the countries, years and sample sizes considered are those listed in Table 1.}
\label{tab:Table2}
\vspace{0.5cm}
\begin{threeparttable}
\resizebox{!}{4.6cm}{\begin{tabular}{c|c|c|c|c|c|c|c|c|c|c|}
\cline{2-11}
&1996&1997&1998&1999&2000&2001&2002&2003&2004&2005\\
\hline
\multicolumn{1}{|c|}{BEL}&0.99$\pm$0.06&0.97$\pm$0.06&0.93$\pm$0.06&0.93$\pm$0.06&0.86$\pm$0.06&0.95$\pm$0.06&0.97$\pm$0.07&0.89$\pm$0.06&0.91$\pm$0.06&0.92$\pm$0.11\\
\multicolumn{1}{|c|}{FIN}&1.25$\pm$0.28&1.45$\pm$0.19&1.49$\pm$0.16&1.40$\pm$0.16&1.21$\pm$0.12&1.26$\pm$0.12&0.94$\pm$0.16&0.99$\pm$0.12&0.99$\pm$0.19&0.83$\pm$0.15\\
\multicolumn{1}{|c|}{FRA}&0.92$\pm$0.10&1.02$\pm$0.11&0.76$\pm$0.07&0.85$\pm$0.07&0.73$\pm$0.05&0.69$\pm$0.04&0.66$\pm$0.03&0.64$\pm$0.03&0.74$\pm$0.05&0.55$\pm$0.08\\
\multicolumn{1}{|c|}{GER}&1.53$\pm$0.13&1.44$\pm$0.09&1.59$\pm$0.09&1.48$\pm$0.09&1.45$\pm$0.11&1.41$\pm$0.10&1.05$\pm$0.18&0.94$\pm$0.13&1.43$\pm$0.08&1.30$\pm$0.22\\
\multicolumn{1}{|c|}{EASTGER}&1.44$\pm$0.16&1.33$\pm$0.22&1.34$\pm$0.20&1.30$\pm$0.19&1.15$\pm$0.16&1.45$\pm$0.25&1.28$\pm$0.14&1.15$\pm$0.11&1.19$\pm$0.14&0.86$\pm$0.17\\
\multicolumn{1}{|c|}{WESTGER}&1.38$\pm$0.15&1.40$\pm$0.09&1.52$\pm$0.14&1.46$\pm$0.10&1.36$\pm$0.11&1.40$\pm$0.12&1.35$\pm$0.10&1.16$\pm$0.12&1.35$\pm$0.11&1.21$\pm$0.26\\
\multicolumn{1}{|c|}{ITA}&1.43$\pm$0.15&1.47$\pm$0.17&1.42$\pm$0.10&1.34$\pm$0.08&1.02$\pm$0.06&1.13$\pm$0.09&0.87$\pm$0.06&1.16$\pm$0.08&1.04$\pm$0.07&1.09$\pm$0.09\\
\multicolumn{1}{|c|}{NORTHITA}&1.37$\pm$0.23&1.38$\pm$0.23&1.49$\pm$0.13&1.31$\pm$0.09&0.96$\pm$0.07&1.19$\pm$0.12&0.83$\pm$0.07&1.38$\pm$0.09&1.12$\pm$0.07&1.11$\pm$0.10\\
\multicolumn{1}{|c|}{SOUTHITA}&1.13$\pm$0.17&1.84$\pm$0.10&1.83$\pm$0.10&1.58$\pm$0.08&1.16$\pm$0.10&1.10$\pm$0.13&0.99$\pm$0.10&1.17$\pm$0.09&1.06$\pm$0.09&1.07$\pm$0.17\\
\multicolumn{1}{|c|}{NET}&1.58$\pm$0.18&1.38$\pm$0.15&1.61$\pm$0.14&0.83$\pm$0.14&0.97$\pm$0.11&0.64$\pm$0.06&0.62$\pm$0.06&0.65$\pm$0.06&0.99$\pm$0.08&0.96$\pm$0.15\\
\multicolumn{1}{|c|}{SPA}&1.10$\pm$0.10&1.19$\pm$0.09&1.17$\pm$0.07&1.04$\pm$0.06&0.97$\pm$0.05&0.90$\pm$0.04&0.85$\pm$0.04&0.78$\pm$0.03&0.74$\pm$0.03&0.98$\pm$0.27\\
\multicolumn{1}{|c|}{SWE}&n.a.&1.17$\pm$0.08&0.98$\pm$0.05&1.05$\pm$0.06&0.97$\pm$0.06&1.00$\pm$0.07&1.05$\pm$0.07&1.04$\pm$0.05&0.97$\pm$0.05&1.12$\pm$0.08\\
\multicolumn{1}{|c|}{UK}&0.99$\pm$0.07&0.97$\pm$0.07&0.93$\pm$0.07&0.96$\pm$0.07&0.93$\pm$0.07&0.96$\pm$0.05&0.97$\pm$0.05&0.96$\pm$0.05&0.93$\pm$0.05&0.94$\pm$0.07\\
\hline
\end{tabular}}
\begin{tablenotes}
\item[\textit{a}]\footnotesize The 95\% confidence intervals around the point estimates are given by $\alpha\pm f_{95\%}\frac{\alpha}{\sqrt{m}}$, where $f_{95\%}$ is the 95\% point of the normal\\
distribution.
\end{tablenotes}
\end{threeparttable}
\end{sidewaystable}
Inspection of the results reveals slight differences among years and countries: for example, for some countries (Belgium, West Germany, Sweden and the United Kingdom) we observe relatively homogeneous entries for the tail indices, while for other countries (Finland, France, East Germany and Spain) the estimate of $\alpha$ has a tendency to decrease in time; exceptions to these temporal patterns are Germany, Italy, North and South Italy and Netherlands, for which the value of $\alpha$ shows a sharp decrease around the beginning of the current decade. However, the time interval under investigation is too short to decide whether these differences are due to major economic and/or political-institutional changes which could have led to a change of the extremal part of the distributions\footnote{For a more in-depth investigation of the tail behaviour, we have also fitted our data to the $\alpha$-stable distribution using the  program \textsc{Stable} \citep{Nolan1997,Nolan1999a,Nolan1999b,Nolan2001}, available from J. P. Nolan's website: \url{academic2.american.edu/~jpnolan}. We noticed that only in a small number of cases the 95\% confidence intervals of the semi-parametric tail index estimates extend to the realm of stable laws, and that in a more limited number of cases the tail index estimates calculated from the stable model are in a somewhat close accordance with the semi-parametric ones. But this is to be expected, since semi-parametric tail index estimation provides a tight fit of the distribution outer parts, whereas the stable law parameters are selected to approximate the entire shape of the empirical distribution \citep{DuMouchel1983,Lux1996}.}.


\subsection{Power-law-tailed distributions in firms' interaction networks}
\citet{DiMatteoAsteGallegati2005} have provided a simple model of technological change through a \textit{social network} of interactions between firms to explain the occurrence of power-law tails in the empirically observed productivity distributions. The general idea behind this work is that a productivity-increasing technological innovation, originally introduced and adopted by a certain firm, can spread over time to other firms by imitation if they interact through a ``scale-free'' type network with degree distribution given by $p\left(k\right)\sim k^{-\left(\alpha+1\right)}$. The model predicts that the aggregate distribution for the productivity of the ensemble of firms is given by a normalized sum of Gaussians with averages distributed according with the connectivity in the network of interactions among firms. Therefore, it is the special structure of the underlying network, having slow decaying tails in its degree distribution, which shapes the aggregate productivity distribution. This theoretical prediction results in good quantitative agreement with the empirical results for the productivity distribution in France and Italy in the years 1996--2001 based on the ``TOP 1.5 million Module'' of \textsc{Amadeus} database\footnote{In the ``TOP 1.5 million Module'', British, French, German, Italian, Russian, Spanish and Ukrainian companies are included if they satisfy at least one of the following criteria: operating revenues bigger than 1.5 million \euro; total assets bigger than 3 million \euro; number of employees bigger than 20. For all other countries, companies are included if their operation revenue is bigger than 1 million \euro, or total assets are bigger than 2 million \euro, or the number of employees is bigger than 15.}.
\par
\begin{figure}[t]
\centering
\includegraphics[width=1.00\textwidth]{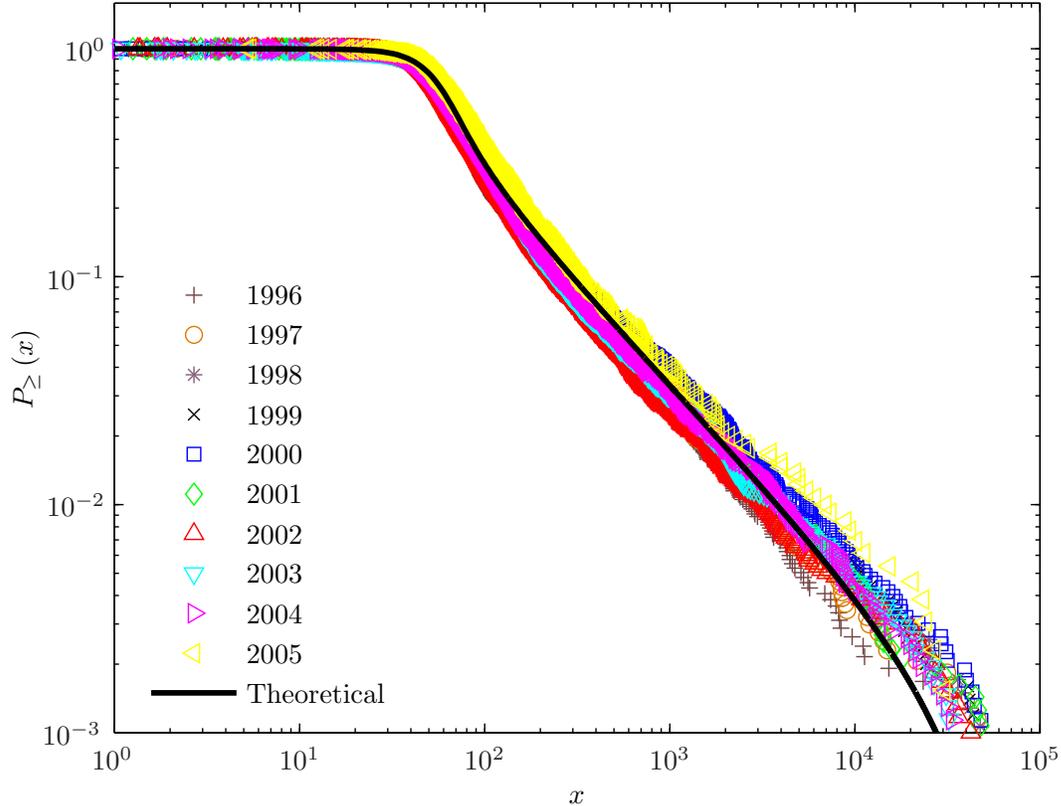}
\caption{Complementary cumulative distributions for the labour productivity in Belgium over the years 1996--2005. The theoretical behaviour (black solid line) is for $\alpha=1.84$, $m=33$, $n=20$, $\sigma=16$ and $\beta=0.5$.}
\label{fig:BEL_final}
\end{figure}
\begin{figure}[t]
\centering
\includegraphics[width=1.00\textwidth]{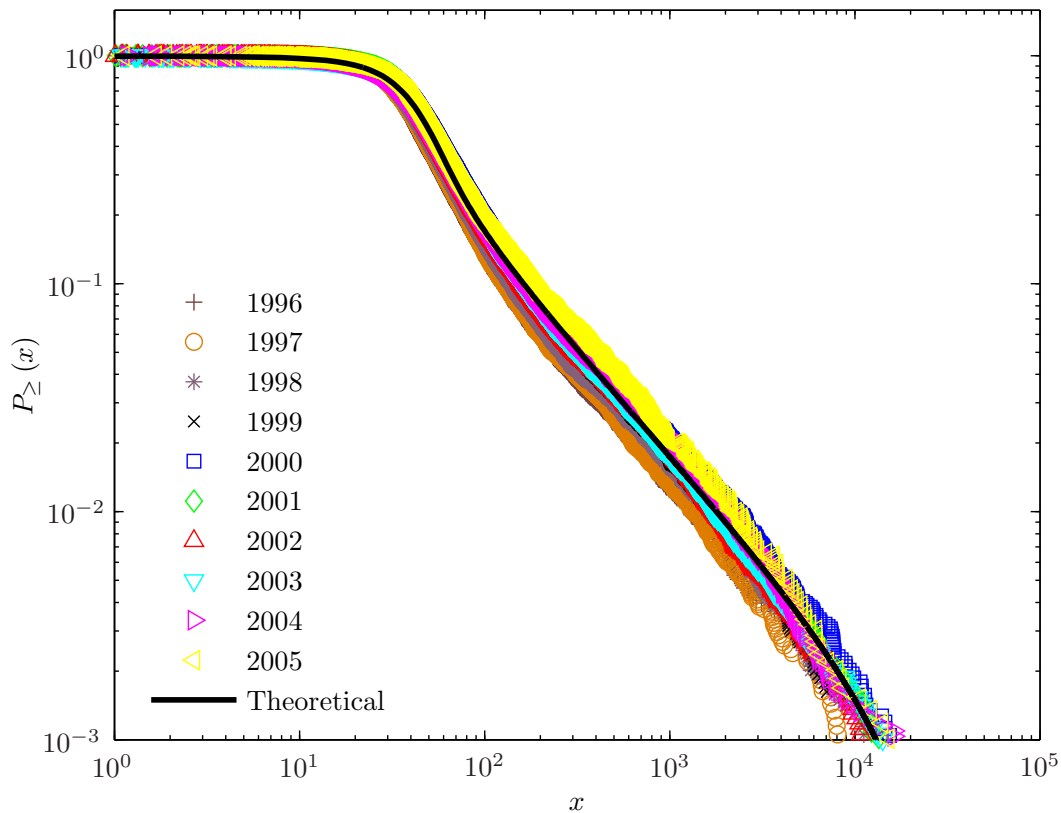}
\caption{Complementary cumulative distributions for the labour productivity in the United Kingdom over the years 1996--2005. The theoretical behaviour (black solid line) is for $\alpha=1.88$, $m=24$, $n=11$, $\sigma=16$ and $\beta=0.6$.}
\label{fig:UK_final}
\end{figure}

Here we extend the analysis to actual empirical evidence coming from our dataset of firms fulfilling the ``TOP 250,000 Module'' inclusion criteria. Figs. \ref{fig:BEL_final} and \ref{fig:UK_final} show the log-log plot of the complementary cumulative distributions of labour productivity corresponding to the years 1996--2005 for two different countries: Belgium and the United Kingdom\footnote{Productivity data have been deflated by using the implicit GDP deflator ($2000=100$) taken from the OECD Statistics Portal (\url{www.oecd.org/statistics/}).}.
We find a quantitatively good agreement by considering an underlying scale-free network with degree distribution given by $p\left(k\right)\propto k^{-\left(\alpha+1\right)}\exp\left(-\beta/k\right)$, averages $k^{\left(1\right)}_{l}=m+z_{l}n$ directly proportional to the number of connections $z_{l}$ that each firm $l$ has in the network, and variance equal to $\sigma$. We note that, although there are several parameters to calibrate, the tail behaviour of the theoretical distribution is controlled only by the power-law exponent $\alpha$, while in the small and medium ranges the other parameters have a larger influence. From our analysis we observe that the theoretical curves (drawn as solid lines) fit well the empirical findings with $\alpha=0.84$, $m=33$, $n=20$, $\sigma=16$ and $\beta=0.5$ for Belgium, and $\alpha=0.88$, $m=24$, $n=11$, $\sigma=16$ and $\beta=0.6$ for the United Kingdom. Very good levels of agreement (not shown here but available upon request) have also been obtained for the other countries considered in our study; the parameters used for the theoretical curves are shown in Table \ref{tab:Table3}.
\begin{table}[t]
\centering
\caption{Model parameters used to draw the theoretical curves for all countries and geographical sub-areas.}
\label{tab:Table3}
\vspace{0.5cm}
\begin{tabular}{c|c|c|c|c|c|}
\cline{2-6}
&$\alpha$&$m$&$n$&$\sigma$&$\beta$\\
\hline
\multicolumn{1}{|c|}{BEL}&0.8&33&20&16&0.5\\
\multicolumn{1}{|c|}{FIN}&1&30&14&12&0.5\\
\multicolumn{1}{|c|}{FRA}&0.7&30&10&12&0.3\\
\multicolumn{1}{|c|}{GER}&1&36&15&18&0.9\\
\multicolumn{1}{|c|}{EASTGER}&1.3&10&6&8&10\\
\multicolumn{1}{|c|}{WESTGER}&1.1&35&18&16&1\\
\multicolumn{1}{|c|}{ITA}&1.1&32&13&18&1.5\\
\multicolumn{1}{|c|}{NORTHITA}&1.2&28&19&16&1\\
\multicolumn{1}{|c|}{SOUTHITA}&1.1&23&19&16&1\\
\multicolumn{1}{|c|}{NET}&0.8&34&14&16&0.6\\
\multicolumn{1}{|c|}{SPA}&0.8&21&13&18&0.2\\
\multicolumn{1}{|c|}{SWE}&0.9&30&7&12&1.5\\
\multicolumn{1}{|c|}{UK}&0.9&24&11&16&0.6\\
\hline
\end{tabular}
\end{table}
Notice that, although there is still matching between the theoretical predictions and the empirical findings, the numerical values we need to theoretically approximate the shape of the East German empirical distributions in an appropriate way are somewhat different from those of the other countries. This might be due to the limited number of entries this geographical area accounts for over the entire period of investigation, which shapes the productivity distributions differently than the others, especially in their outer parts.


\section{Social network, social capital and economic performance}
\label{sec:SocialNetworkSocialCapitalAndEconomicPerformance}
A huge literature focuses on the relationship among \textit{social capital} and productivity of economic units or organizations \citetext{see \citealp{CohenPrusak2001}, among others}. In particular, some authors \citetext{\textit{e.g.} \citealp{Fukuyama2000}} tend to put emphasis on qualitative aspects of the relationship network-capital, drawing the attention to the capability of social capital within developed societies of linking heterogeneous social networks and improving communication and information flows. Along these lines, our aim here is to investigate whether social capital plays a role in the transfer of knowledge, information, and technology through the social network of firms in a country \citetext{on this point see in particular \citealp{Ahuja2000}}. The basic hypothesis is that a higher level of social capital improves the efficacy of social network linkages, favoring and strengthening connections among agents, and lowering costs and time of communication \citep{Coleman1988}. In terms of the present study, a more effective social network reduces the relative distances among firms' productivity levels, since innovation and technological information flow more rapidly and with lower costs \citep{Granovetter1985}. Firms' productivity, therefore, results more evenly distributed, and the power-law exponents increase. The verification of this hypothesis introduces an original way to investigate the relation among social capital of a country and economic performance at aggregate level, since countries with different levels of social capital should display as well a different power-law exponent in labour productivity distribution.
\par
By social capital we mean the `features of social life-networks, norms, and trust, that enable participants to act together more effectively to pursue shared objectives' \citep[pp.~664--665]{Putnam1995}. This definition supports the choice of performing a country-level analysis, since a nation is supposed to represent a homogeneous sample as regards social network and institutional aspects. The concept of social capital was firstly introduced in sociology with reference to groups or communities. The extension of the concept at the country level, operated by political scientists, has been initially subject to critics, especially as regards measurement and distinction among human and social capital \citep{Solow1995}. In more recent years, some of the cited studies performed at country \citep{Coleman1988} and sub-country level \citetext{\textit{e.g.} \citealp{DiGiacintoNuzzo2006}} demonstrated the usefulness of the concept of social capital, in particular for investigating social network features. Besides, \citet{Putnam2000} stresses the relevance of social capital in improving the performance of individuals, since it puts them in a connected network. Indeed, social networks are often identified by specialized literature as the ``structure'' of social capital \citetext{\textit{e.g.} \citealp{Burt2000}}, concept that is well specified by \citet[p.~249]{Bourdieu1996}: `The volume of social capital possessed by a given agent [\ldots] depends on the size of the network of connections that he can effectively mobilize'. In order to obtain a synthetic indicator, empirical analyses usually adopt international surveys. Along the lines of most of these works, we employ the \textit{World Values Survey} (\textit{WVS}), a data source designed to enable a cross-national/cross-cultural comparison of values and norms in a wide variety of areas, and to monitor changes in values and attitudes in societies all over the world\footnote{To date, the World Values Survey has carried out four waves (1981--1984, 1989--1993, 1994--1999, and 1999--2004) of national surveys representative of the values and beliefs in more than 80 countries on all six inhabited continents. The data are available for free download from the project website: \url{http://www.worldvaluessurvey.org/}.}. In particular, we refer to the latest available wave of the WVS by adopting as a social capital measure the \textit{trust}, quantified by the percentage of interviewed people who agree to the assertion that ``most people can be trusted''\footnote{The exact question in the WVS is: ``Generally speaking, would you say that most people can be trusted, or that you can't be too careful in dealing with people?''.}. According to \citet{KnackKeefer1997} and \citet{Sabatini2006}, among others, this quantity is likely to be deeply related with economic and productivity performances\footnote{In order to avoid biases due to the oversampling of certain categories of people interviewed, all the answers to these questions have been pondered by the weights provided in the survey itself.}. Moreover, the use of this proxy permits to avoid \citeauthor{Portes1998}' \citeyearpar{Portes1998} critic, according to which the isolation of social capital's definition from its effects would be ambiguous and, with particular reference to trust, could be reduced to the result of the effectiveness of legal enforcement in a country's system \citep{GuisoSapienzaZingales2004}. This linkage among social capital and legal enforcement in the WVS is better captured by the variable \textit{civic}, which concerns a series of social behaviors (such as ``avoiding a fare on public transport'') that can be never, partially or totally justified by the interviewed persons \citetext{see \citealp{KnackKeefer1997}, for a more detailed explanation and an investigation of the relationship among civic and trust}. Nevertheless, the measure of interpersonal trust reported in the WVS appears to be consistent with its definition as an equilibrium outcome of a society where non-legal mechanisms force people to behave cooperatively \citep{Coleman1990}.
\par
\begin{figure}[t]
\centering
\includegraphics[width=1.00\textwidth]{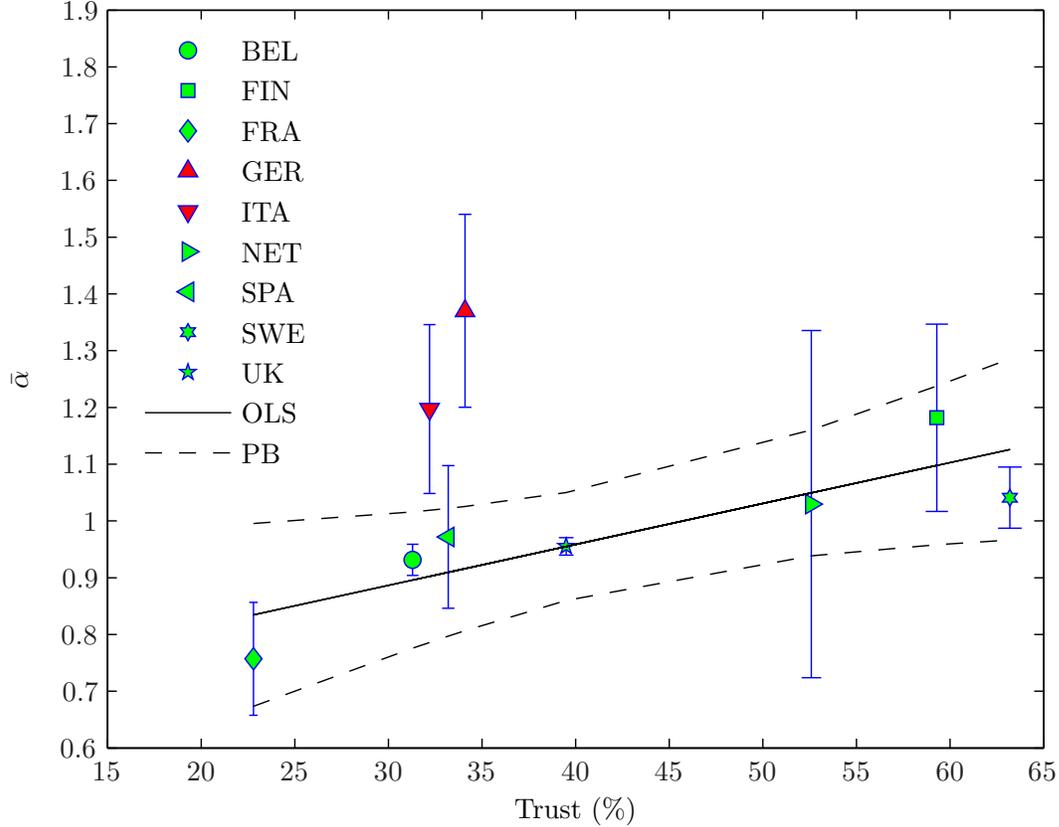}
\caption{Error bar plot of the average tail index estimate against the WVS-based trust measure. The length of each error bar equals two times the standard error of the mean. The black solid line is the ordinary least squares (OLS) fit to the data with 95\% prediction bounds (PB).}
\label{fig:WVS_final}
\end{figure}

On Fig. \ref{fig:WVS_final} we plot $\bar{\alpha}$, the average of the power-law exponent estimates for each country over the period under investigation, as a function of the level of trust; the corresponding values are shown in Table \ref{tab:Table4}\footnote{Due to the reduced number of observations, which might bias the results for some countries, we exclude the tail index estimates for 2005 from the computation of the mean for countries with less than 1000 observations in that year (namely, Germany, West Germany, South Italy, Netherlands and Spain). As regards East Germany, the mean is computed considering all values, since anyway the number of observations is always less than 1000. Notice that the choice to use the average value of the estimates permits to smooth temporary variations (that in some countries\textemdash \textit{e.g.} Germany, Italy and Netherlands\textemdash are not negligible), and it is likely to be more appropriate to enable comparison with the wave of WVS data we use, since this data collection was undertaken in the central years of the period under analysis for firms.}.
\begin{table}[t]
\centering
\caption{Temporal averages of the power-law exponent estimates ($\bar{\alpha}$) over the period under investigation and percentage level of trust (WVS, 1999-2004 wave) for each country and geographical sub-area considered in the study. Also shown is the estimated standard error of the mean.}
\label{tab:Table4}
\vspace{0.5cm}
\begin{threeparttable}
\begin{tabular}{c|c|c|}
\cline{2-3}
&\textbf{$\bar{\alpha}$}&Trust (\%)\\
\hline
\multicolumn{1}{|c|}{BEL}&0.93$\pm$0.03&31.30\\
\multicolumn{1}{|c|}{FRA}&1.18$\pm$0.17&59.30\\
\multicolumn{1}{|c|}{FIN}&0.76$\pm$0.10&22.80\\
\multicolumn{1}{|c|}{GER}&1.37$\pm$0.17\tnote{\textit{a}}&34.10\\
\multicolumn{1}{|c|}{EASTGER}&1.29$\pm$0.09&48.30\\
\multicolumn{1}{|c|}{WESTGER}&1.36$\pm$0.08\tnote{\textit{a}}&40.70\\
\multicolumn{1}{|c|}{ITA}&1.20$\pm$0.15&32.20\\
\multicolumn{1}{|c|}{NORTHITA}&1.23$\pm$0.17&n.a.\\
\multicolumn{1}{|c|}{SOUTHITA}&1.32$\pm$0.26\tnote{\textit{a}}&n.a.\\
\multicolumn{1}{|c|}{NET}&1.03$\pm$0.31\tnote{\textit{a}}&52.60\\
\multicolumn{1}{|c|}{SPA}&0.97$\pm$0.13\tnote{\textit{a}}&33.20\\
\multicolumn{1}{|c|}{SWE}&1.04$\pm$0.05\tnote{\textit{b}}&63.20\\
\multicolumn{1}{|c|}{UK}&0.96$\pm$0.02&39.50\\
\hline
\end{tabular}
\begin{tablenotes}
\item[\textit{a}]\footnotesize Excluding 2005.
\item[\textit{b}]\footnotesize Excluding 1996.
\end{tablenotes}
\end{threeparttable}
\end{table}
By observing the graph, one can notice that a tendency toward a positive trend seems to emerge between the average value of the estimates of power-law exponents and the level of trust. However, given the low number of cases included in the analysis, it is not possible to infer any further conclusions. Nonetheless, some deviations from this trend are present. In particular, the calculated value of the linear (Pearson's) correlation coefficient between these two variables is 0.28, with an estimated $p\text{-value}$ for testing the hypothesis of no correlation equal to 0.46; however, once Germany and Italy have been excluded from the calculation, the estimated correlation coefficient and $p\text{-value}$ are 0.86 and 0.01, respectively. These results are confirmed if one uses Kendall's $\tau$ and Spearman's $\rho$ as more general and robust measures of dependence, obtaining $\tau=0.39$ ($p\text{-value}=0.18$) and $\rho=0.47$ ($p\text{-value}=0.21$) when the two countries are included in the analysis, and $\tau=0.81$ ($p\text{-value}=0.01$) and $\rho=0.93$ ($p\text{-value}=0.01$) when they are not. The positive but significant (at the 5\% significance level) correlation only once Germany and Italy are excluded from the computation points to an \textit{outlying} behaviour of these countries, which indeed reveal an average value of the power-law exponent significantly higher compared to the other countries. A possible explanation of this behaviour involves the particular heterogeneity within each country. The aggregates of these two countries are actually the sum of two different social networks and economic systems: East and West for Germany, North and South for Italy\footnote{See \citet{Vecernik2003} for Germany and \citet{DiGiacintoNuzzo2006} for Italy.}. The average values of the tail index estimations for the above-mentioned levels of geographical disaggregation is shown in the second column of Table \ref{tab:Table4}. As regards Germany, the value of trust is somewhat bigger in the Eastern part (48.3\% against 40.7\% of the West), but it should be noted that at the beginning of the period under observation for firms the percentages of trust were 24.3 for the East and 39.9 for the West, respectively (1997 WVS data)\footnote{These results do not differ greatly from the values of immediate pre-unification period. Indeed, East and West Germany's 1990 WVS percent levels of trust equalled to 20.1 and 31.1, respectively.}. During this time, firms in East Germany were catching up Western ones: the aggregate labour productivity of East Germany (as a percentage of the West Germany's level) progresses from 45\% in 1990 to approximatively 70\% in 2002 \citep{Uhlig2006}; simultaneously, the power-law exponent estimates of the Eastern firms' productivity distribution results lower at the end of the period of observation with respect to the beginning, while they remain substantially stable in West Germany\footnote{A word of caution is needed here due to the low number of observations for East Germany.}. If considered togheter, these matters suggest that the improvement in Eastern workers' productivity has been accompanied by a relevant integration of the different social networks and an increase of the differences between firms in the initially disadvantaged area. Therefore, a Schumpeterian mechanism seems to be at work here: not all firms took advantage from the new body of technologies and information available. In particular, the augmented level of social trust did not determine a generalized improvement in firms' productivity due to the massive migration of workers towards the West and, consequently, the difficulties for Eastern firms in hiring skilled workers. According to \citet{Cooper1999}\footnote{But see also \citet{RosenfeldFranzGuntherHeimpoldKawkaKronthalerBarkholz2004} on the related question of ``clusterization'' of Eastern firms.}, this networking problem is at the root of the slowdown in the catching-up process observed after 2002. In other words, over a certain starting threshold of heterogeneity, even a remarkable improvement in social capital has limited or no effect on the network structure, the communication among weakly connected points being problematic (the ``structural holes'' proposed by \citeauthor{Burt2000}). As regards Italy, given the unavailability of geographical sub-area survey data, no definitive conclusions can be drawn from the disaggregation analysis, even though the average level of power-law exponent for Northern firms is slightly lower.


\section{Conclusions}
\label{sec:Conclusions}
In this work we have detected the emergence of power-law tails in labour productivity distributions for 9 European countries and different time periods. We have modeled the empirical labour productivity distributions with the model introduced by \citet{DiMatteoAsteGallegati2005}, and compared its outcomes with the empirical power-law exponents estimated by means of \citeauthor{ClementiDiMatteoGallegati2006}'s \citeyearpar{ClementiDiMatteoGallegati2006} algorithm. The model has been validated for all cases, confirming that power-law tails can emerge from scale-free contact-networks. Moreover, we have investigated the relationship between productivity distribution features and social trust, evidencing a tendency toward a positive relationship between the mean values of the power-law exponents of labour productivity and the level of trust. However, the data appear scattered and, because of the reduced number of points, it is not possible to draw a definitive conclusion.


\begin{ack}
Many thanks to Tomaso Aste and an anonymous referee for helpful comments and suggestions. T. Di Matteo wishes to thank the partial support by ARC Discovery Projects: DP03440044 (2003) and DP0558183 (2005), and COST P10 ``Physics of Risk'' project (2003).
\end{ack}


\bibliography{Bibliography}
\bibliographystyle{apacite}
\end{document}